\documentclass[]{aa}
\usepackage{natbib}
\usepackage{graphicx}
\usepackage{wasysym}
\usepackage{txfonts}
\def\ACO{\mbox{A$_{\rm CO}$}}

\def\AV{\mbox{A$_{\rm V}$}}

\def\nH2{\mbox{${\rm n}_\HH}$}

\def\pccc{~{\rm cm}^{-3}} 
 
\def\pcc {~{\rm cm}^{-2}}
\def\Tstar {\mbox{${\rm T}_{\rm r}^*$}}
\def\Tsub#1 {\mbox{${\rm T}_{\rm #1}$}}
\def\TK  {\Tsub K }

\def\fvol{\mbox{f$_{\rm vol}$}}

 \def\arcmin{\mbox{$^{\prime}$}}

\def\degr{$^{\rm o}$}
\def\p{\mbox{$^+$}}

\def\rmR{\mbox{${\rm R}$}}
\def\rmv{\mbox{${\rm v}$}}
\def\rmn{\mbox{${\rm n}$}}
\def\sil{\mbox{$\sin{l}$}}
\def\col{\mbox{$\cos{l}$}}
\def\si2l{\mbox{$\sin^2{l}$}}                                                               
\def\co2l{\mbox{$\cos^2{l}$}}                                                                 
\def\AV{\mbox{A$_{\rm V}$}}

\def\cch{\mbox{C$_2$H}}
  
\def\hhco{\mbox{H$_2$CO}}
\def\h13cop{\mbox{{H$^{13}$CO\p}}}

\def\C3H{\mbox{C$_3$H}}
\def\c3h2{\mbox{C$_3$H$_2$}}
\def\cc3h2{\mbox{{\it c}-C$_3$H$_2$}}

\def\R0{\mbox{R$_0$}}
\def\G0{\mbox{G$_0$}} 
  \def\kpc{\rm kpc}
  
\def\ddeg{{}^\circ\kern-.1em} 
 
\def\thsun{{\Theta_0}}
\def\kms{\mbox{km\,s$^{-1}$}}

\def\E#1 {$10^{#1}$}
\def\E#1 {E{#1}}
\def\P#1,{$\nH2\TK~=~#1\times~10^4\pccc$~K}
\def\ec#1,#2,#3,{#1\,(#2)\E{#3}}

\def\H3{\mbox{H$_3$}}

\def\ammon{\mbox{N\H3} }

\def\RH2{\mbox{R$_{\rm G}$}}
\def\g13{\mbox{g$_{13}$}} 

\def\cc3h{\mbox{{\it c}-\C3H}}
\def\lc3h{\mbox{{\it l}-\C3H}}

\newcommand{\emm}[1]{\ensuremath{#1}}   
\newcommand{\emr}[1]{\emm{\mathrm{#1}}} 

\newcommand{\abs}[1]{\emm{\left|  #1 \right| }} 

\newcommand{\hcop}{\emr{HCO^+}} 
 
\newcommand{\HH}{\emr{H_2}}
\newcommand{\cotw}{\emr{^{12}CO}}
\renewcommand{\coth}{\emr{^{13}CO}}
\newcommand{\coei}{\emr{C^{18}O}}

\newcommand{\X}[1]{\emm{X_\emr{#1}}}

\newcommand{\XCO}{\X{CO}}

\newcommand{\W}[1]{\emm{{\rm W}_\emr{#1}}}
\newcommand{\WCO}{\W{CO}}

\title{Molecular gas in absorption and emission along the line of sight to W31C G10.62-0.38}

\author{H. S. Liszt\inst{1} \& M. Gerin\inst{2}}
\institute{National Radio Astronomy Observatory,
           520 Edgemont Road,
           Charlottesville, VA,
           USA 22903-2475
\and       LERMA, Observatoire de Paris, 
           PSL Research University, CNRS, Sorbonne
           Universit\'es, UPMC Univ. Paris 06, 
           \'Ecole normale sup\'erieure,  F-75005 Paris, 
           France
}

\begin{document}
\date{received \today}%
\offprints{H. S. Liszt}%
\mail{hliszt@nrao.edu}%
%
\abstract
{The sightline to W31C G10.62-0.38 was extensively observed in absorption
 under the PRISMAS program on Herschel.}
{To relate absorbing material to the older 
  view of Galactic molecules gained from CO emission.}
{We used the ARO 12m antenna to observe emission from the J=1-0 lines of
 carbon monoxide, \hcop\ and HNC and the J=2-1 line of CS toward and around 
the continuum peak used for absorption studies and we compare them with
  CH, HNC, C\p\ and other absorption spectra from PRISMAS.
  We develop a kinematic analysis that allows a continuous description
  of the spectral properties and relates them to viewing geometry in the Galaxy.}
{As for  CH, HF, C\p, \hcop\ and other species observed in absorption, mm-wave 
 emission in CO, \hcop, HNC and CS is continuous over the full velocity range 
  expected for material between the Sun and W31 4.95 kpc away.  CO emission is 
 much stronger than average in the Galactic molecular ring and  
 the mean \HH\ density derived from CH, $4 \pccc \la$ 2$<$n(\HH)$>$ $\la 10 \pccc$ 
  at 4 $\la$ R $\la$ 6.4 kpc, is similarly elevated.  The CO-\HH\
 conversion factor falls in a narrow range 
 \XCO\ $= 1-2\times10^{20}~\HH\ \pcc~({\rm K}-\kms)^{-1}$ 
 if the emitting gas is mostly on the near side of the sub-central point, 
 as we suggest.  The brightnesses of \hcop, HNC, and CS are comparable 
 (0.83\%, 0.51\% and 1.1\% respectively relative to CO) and have
  no variation in galactocentric radius with respect to CO.  
 Comparison of the profile-averaged \hcop\ emission brightness and optical depth 
 implies local densities n(H) $\approx 135\pm25\pccc$ with most of excitation 
 of \hcop\ from electrons.  At such density, a consistent picture of the 
\HH-bearing gas, accounting  also for  the CO emission, has a volume filling 
factor 3\% and a 5 pc clump or cloud size.
}
{}


\authorrunning{Liszt and Gerin W31} \titlerunning{W31C}

\maketitle{}

%

\section{Introduction}

The earliest perspectives on the large-scale distribution of molecules in
the disk of the Milky Way were provided by cm-wave surveys of OH \citep{Gos67}
and \hhco\ \citep{WhiGar70} seen in absorption against bright H II regions in the
Galactic plane.  This work extended our view of molecular gas far beyond the
local clouds sampled in optical absorption, but was restricted to relatively
few directions at low latitude and could not  provide such basic quantities
as the Galactic scale height and overall radial distribution of the 
molecular gas, or the size and density of the host gas clouds.  

Absorption line surveys were  overtaken by studies of mm-wave CO emission 
\citep{ScoSol75,BurGor78,CleSan+88,DamHar+01} whose ubiquity and ease of observation
allowed discovery of GMCs (giant molecular clouds) in the disk as well
as the Galactic molecular ring hosting the bulk of the molecular gas. 
Maps of carbon monoxide emission have come to define the Galactic distribution of 
molecular gas.  

Nonetheless, molecular emission studies are subject to their own limitations. Emission
from the J=1-0 line of CO is notoriously insensitive to the gas number density 
\citep{LisPet+10} and Galactic plane surveys of emission from species that are
supposedly sensitive to density such as \hcop\ and HCN \citep{Lis95,HelBli97}
have beeen extremely limited.  They show little or no variation with respect 
to CO across the Galactic disk, which is surprising given the variation of 
the mean molecular density.  In general, emission studies are also bedeviled 
by confusion between dense and diffuse gas, together with uncertainty in the 
volume filling factor.

Turning the tables, so to speak, sub-mm studies of newly-available atomic and 
molecular transitions seen in absorption using the HIFI instrument on the Herschel 
satellite have
recently provided new insights into the physical properties of molecular
gas in the Galactic disk.  Observations of hydride progressions \citep{GerLev+12}
like OH\p, O\HH\p\ and O\H3\p\ \citep{GerdeL+10} or NH, N\HH\ and \ammon\ 
\citep{PerdeL+12} are sensitive to the presence of gas over a wide range of \HH\ 
fraction.

\begin{figure*}
\includegraphics[height=10cm]{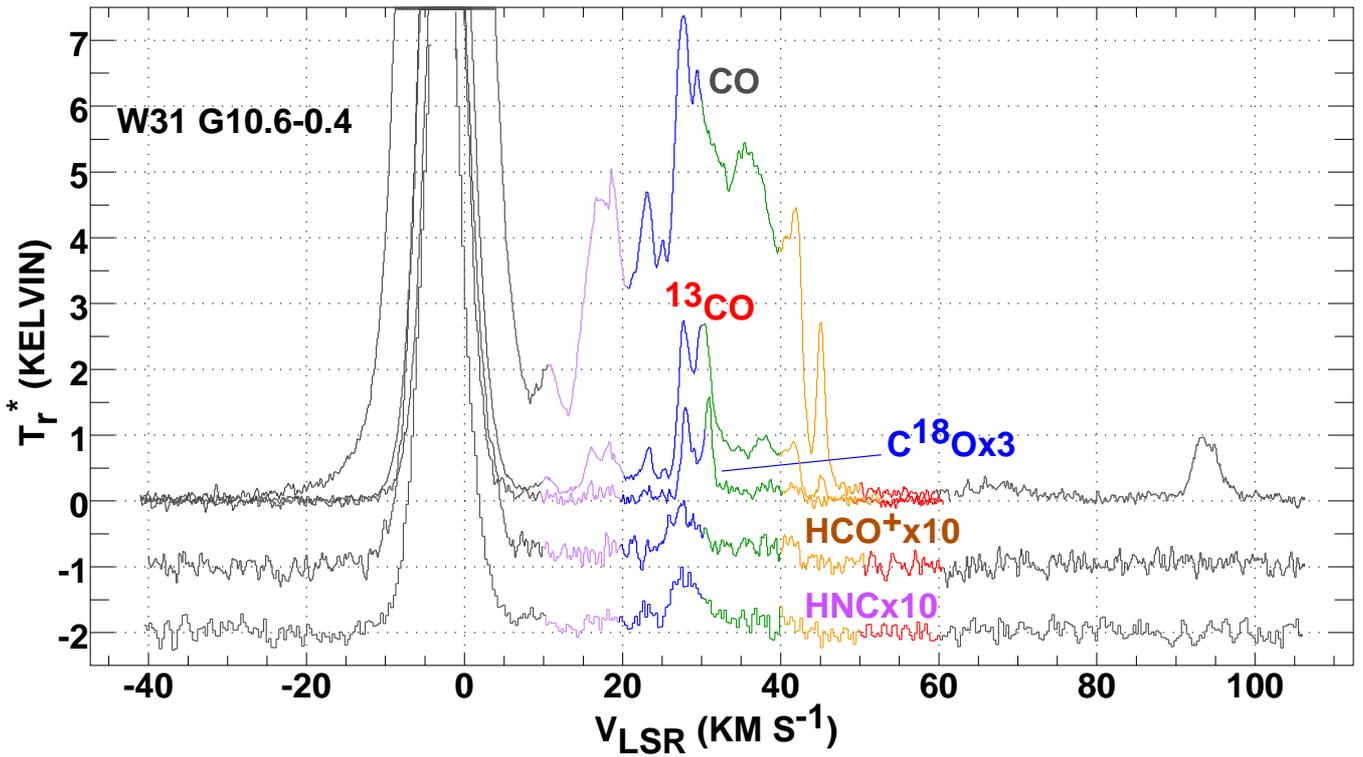}
  \caption[]{Emission spectra observed at the ARO 12m telescope in 2011.  The \hcop\
and HNC spectra are averages of four pointings displaced $\pm$1.4\arcmin\ to the East and
North, to avoid absorption against the strongest continuum.  Spectra are scaled as
indicated.  The spectra are color coded in 10 \kms\ intervals to aid in understanding
of the velocity-distance transformations.}
\end{figure*}

\begin{figure*}
\includegraphics[height=10.2cm]{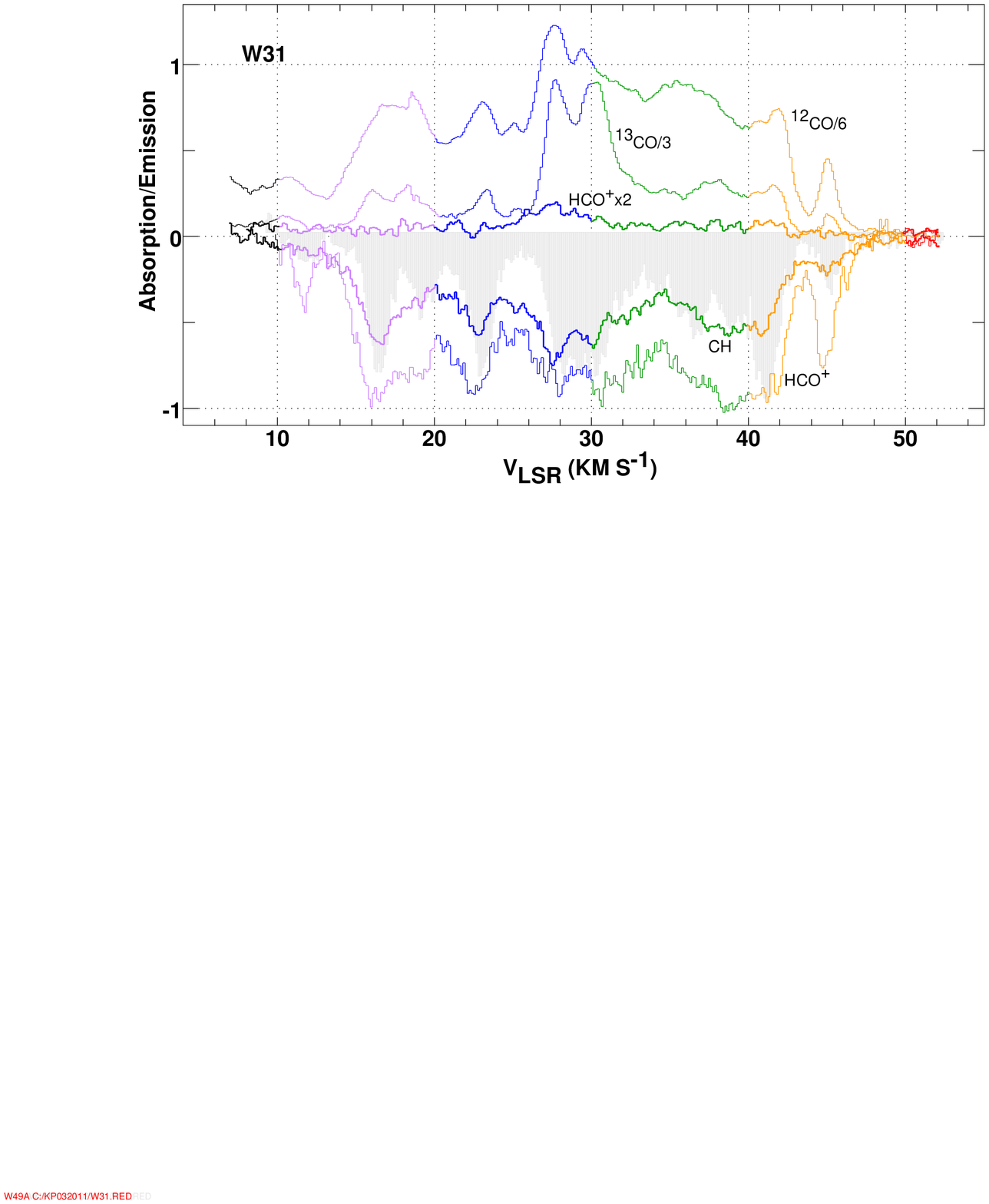}
  \caption[]{Emission and absorption spectra.  The color coding is as in Fig. 1.  Spectra
have been scaled as indicated. The CH and \hcop\ profiles are shown slightly bolder to
remove some possible confusion with other spectra. The absorption spectrum of HNC is
shown as a faint gray underlay.}
\end{figure*}

Here we take a hybrid approach comparing new emission measurements at arcminute 
spatial resolution from the ARO 12m telescope with the wealth of absorption line 
data that was acquired toward W31C G10.62-0.38 in the course of the 
PRISMAS project on Herschel. W31C is the most strongly absorbed  of all the sources
in the PRISMAS sample and was explicitly targeted for its high intervening
molecular column density.  Along this line of sight the Sagittarius, Scutum-
Centaurus and Norma spiral arms all run across the line of sight, as illustrated 
in Figure 6 of  \cite{SanRei+14}, who fixed the distance to W31C as 
$4.95\pm0.5$ kpc.  The viewing geometry  provides a high gas column density, 
having a high overall molecular fraction, projected onto a somewhat narrow 
velocity range 0 - 50 \kms\ because of the low Galactic longitude and
relative proximity of the background target.

The plan of this work is as follows.  The new and 
existing observational material is described in Sect. 2 and discussed in Sect. 3.  
In Sect. 4 we describe a formalism to interpret spectra in a manner that is 
appropriate to the continuous nature of the emission
and absorption profiles seen toward W31C,  in order to derive the variation with 
galactocentric radius of the CO emission, mean \HH\ density and CO-\HH\ conversion 
factor.  This is done in Sect. 5 where we also derive other quantities such as 
the volume filling factor, cloud  size and mean free path in the diffuse molecular 
gas.  Cloud chemical abundances and Galactic gradients in the chemistry are
discussed in Sect. 6 and  Sect. 7 is a summary. 

\section{Observational material discussed here}

\subsection{New mm-wave emission measurements from the ARO 12m}

We observed W31C G10.62-0.38 in ALMA Band 3 (86 - 116 GHz) at the ARO 12m telescope 
in 2011 March.  We took profiles of the J=1-0 lines of \cotw\ (CO), \coth, and 
\coei\ toward the continuum peak and observed the J=1-0 lines of \hcop\ and HNC, and
the J=2-1 line of CS, at positions displaced 1.4\arcmin\ in the four cardinal
directions because spectra taken toward the continuum showed clear absorption.
The \hcop, HNC and CS emission profiles shown here are averages of these four spectra.

All the ARO data were taken by position-switching to a reference position well above
the Galactic plane, whose spectrum (showing weak CO emission near 0-velocity) was
determined by position-switching against a yet more distant reference.  The data
were taken with an autocorrelator resolution of 97.7 kHz (velocity resolution
0.308 \kms\ at 100 GHz) and channel spacing of 48.8 kHz.  Linear baselines
were fit to emission free regions at v $< -30$ \kms\ and v $> 100$ \kms\ for the
carbon monoxide lines or at v $>$ 60 \kms\ for the other species.

The brightness scale of ARO 12m observations is \Tstar.   At 115.3 GHz,
the nominal efficiency factor needed to put such observations on the 
main-beam scale is $(0.84\pm0.07)$
\footnote{see http://www.cv.nrao.edu/~jmangum/12meter/efficiencies.html}.  
However, the emission observed here
is quite extended, even on 1\degr\ scales, and no correction
was applied to the observed 12m brightness.
The rms noise of the spectra shown in Fig. 1 is 0.054 K for
CO and 0.026 K for \coth\ and \coei.  The rms of the HNC and \hcop\ spectra 
is 0.012 K and that of CS slightly larger, 0.016 K.

We also mapped J=1-0 \coth\ emission on-the-fly (OTF) in a 30\arcmin\ field 
around the continuum peak using 250 kHz filters having a velocity resolution 
and channel spacing  0.681 \kms\ at 110.2 GHz.  The same reference position
was used as for the pointed observations toward/around the continuum peak and
the off-spectrum was added back into the cube.  The typical rms in a single
profile is 0.26 K.

All velocities used here are with respect to the kinematic definition of the Local
Standard of Rest 
and labeled LSR. The high spectral resolution pointed emission profiles toward 
and around W31 are shown in Figs. 1, 2 and A.1.  Results of the \coth\ OTF mapping 
are shown in Figs. 3 and C.1.  Table 1 gives integrated intensities and ratios 
with respect to CO for the pointed observations shown in Fig. 1, 2 
and A.1 (for CS).

\begin{table}
\caption[]{Properties of emission spectra}
{
\small
\begin{tabular}{lcccc}
\hline
Species & rms  & $\int \Tstar {\rm dv} $ & $\int \Tstar {\rm dv}$/\WCO  
 &$\int \Tstar {\rm dv}$/\WCO \\
        & K    &  K-\kms                  & integrated  & point-point     \\
\hline
CO    & 0.054  & 145.64(0.110)&&       \\
\coth & 0.028  & 28.39(0.066) & 0.1950 & 0.1694(0.0805)  \\
\coei & 0.024 & 2.82(0.054)  & 0.0194 & 0.0175(0.0275)       \\
\hcop & 0.012 & 1.16(0.030)  & 0.0080 & 0.0082(0.0046) \\
HNC   & 0.012 & 0.76(0.030)  & 0.0052 & 0.0050(0.0043)\\
CS    & 0.016 & 1.59(0.043)  & 0.0109 & 0.0101(0.0163)\\
\hline
\end{tabular}}
\\
\end{table}

\subsection{Other, existing, data}

As discussed below, we used the 532 GHz CH profile of \cite{GerdeL+10a}
to derive the molecular hydrogen abundance assuming a value for 
X(CH) = N(CH)/N(\HH).  This spectrum is shown in
Fig. 2 with the \hcop\ absorption spectrum of \cite{GodFal+10}
and in Fig. 3 with the HNC absorption spectrum of \cite{GodFal+10}.
(the HNC absorption spectrum is also shown in Fig. 2).  
The velocity resolution of the CH profile is 0.067 \kms\ and that for 
HNC is 0.26 \kms.  The C II absorption profile of \cite{GerRua+15} is not shown 
here but is used in Fig. 6 to derive the total density n(H).

\begin{figure}
\includegraphics[height=7.5cm]{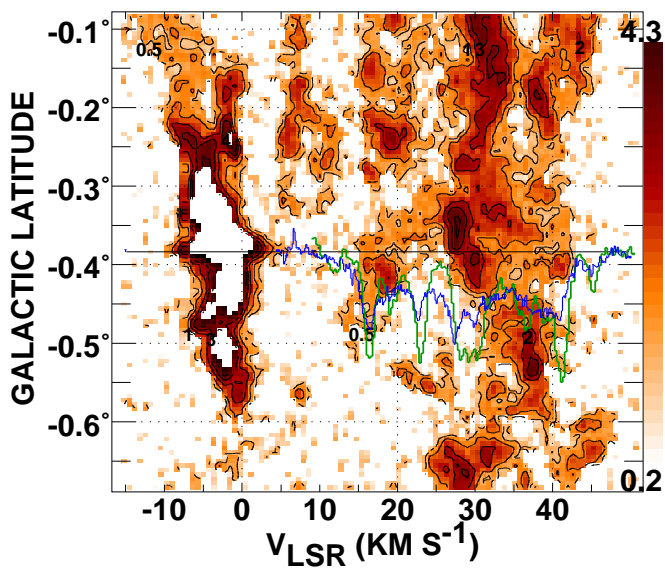}
  \caption[]{Latitude-velocity diagram of \coth\ emission through the position
  of W31C G10.62-0.38 at 1\arcmin\ resolution from the ARO 12m telescope.  
 The \coth\ brightness scale is indicated at right, running from 0.2 to 4.3 K.
  The HNC absorption profile of \cite{GodFal+10} is shown overlaid in green with its 
 zero-level at the latitude of W31 and the CH absorption profile is overlaid in blue
(see also Fig. 2).}
\end{figure}

\section{Observational results}

New ARO emission spectra are shown in Fig. 1.  Emission associated with W31
is centered just below zero velocity and extends to about $+10$ \kms, narrowing
the range of galactocentric radius about which information may be derived.
The bulk of the Galactic emission is seen at 10 - 50 \kms\ but a wing of
weak CO emission extends to 100 \kms.  A pair of narrow features is prominent
in \coth\ and \coei\ at v = 28 - 30 \kms\ and these high column density features
presumably arise in the most fully molecularized material along the line of sight.  

Unlike most other inner-galaxy CO profiles, the CO emission generally does
not break up into discrete features attributable to clouds,  separated by 
velocity intervals where emission is weak or absent.  \coth\ emission is 
also continuous
over the 10 - 45 \kms\ range and even \coei\ is detected over most of
the interval.  Profiles of \hcop\ and HNC emission are shown in Fig. 1 and
a closeup comparison including the profile of CS J=2-1 is shown in Fig. A.1.  
The effective beamwidth of the profiles for \hcop, HNC and CS is broadened
by averaging over positions displaced 1.4\arcmin\ from the continuum but the
broad emission out to 42 \kms\ is clearly detected at all positions  
even in these supposedly density-sensitive species that are so much harder 
to excite than CO.

Fig. 2 shows a comparison of emission and absorption profiles.  With the 
exception of the 45 \kms\ feature that is somewhat indistinct in CH (and much more 
marked in \hcop), CO emission closely mirrors the absorption profiles.  This is 
one indication that the emitting gas lies preponderantly on the
near side of the Galaxy; another is that the CO emission is undetected at
adjacent higher velocities just outside the kinematic span of the absorption profiles.  
Perhaps most  importantly, no gas seen in absorption is missed in the emission profiles 
of CO and \coth\ and perhaps even in \coei. 

\begin{figure}
\includegraphics[height=5.25cm]{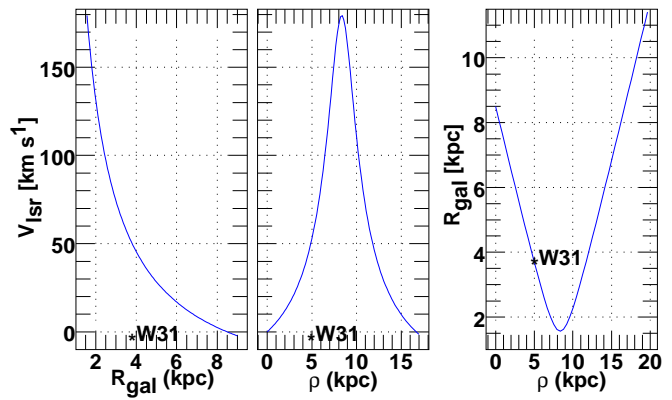}
  \caption[]{Velocity and distance transformations toward W31.  Left and middle:  
LSR velocity vs. galactocentric distance R and line of sight distance $\rho$
 for a flat rotation curve with \R0\ = 8.5 kpc, $\Theta$(R) =  220 \kms.  Right: 
the galactocentric distance-line of sight distance relationship for \R0\ = 8.5 
kpc.  The locus of W31 is noted in all panels, based on the 4.95 kpc distance measured 
by \cite{SanRei+14}.}
\end{figure}

Fig. 3 shows a latitude cut through the \coth\ emission datacube, with the position
of W31 indicated and the absorption profiles of CH and HNC superposed with
their zero-levels there.  The signal/noise ratio of this lower-resolution 
\coth\ dataset is substantially poorer than that of the pointed observations
illustrated in Figs 1-2 (the typical rms is 0.26 K) . The narrow features at 28 - 30 \kms\
toward W31 are slightly separated in space but both lie in a band of widely-distributed
strong emission.  The spatial distribution illustrated in Fig. A1 shows that this
gas lies in the outskirts of a GMC having very strong emission to the North and West.

The latitude distribution of the  brightest gas in Fig. 3, at v $>$ 25 \kms, is nearly 
symmetrical about the latitude of W31, suggesting that it is relatively nearby
on the near side of the Galaxy.  By contrast, the emission at 16 - 21 \kms\ is
noticeably stronger nearer b = 0\degr.  It coincides with a local minimum in the
absorption and is more likely to have a larger contribution from far-side gas.

\section{Kinematic analysis of the line of sight toward W31}

\subsection{Galactic rotation and the velocity-galactocentric radius transformation}

Previous analyses of material along individual sightlines like that toward
W31 have treated line profiles using  a set of velocity intervals
determined by the observed spectral features.  Here we develop a formalism that
treats line profiles more continuously and provides  an innate relationship
to line of sight distance and the Galactic-scale kinematic effects that shape the
profiles.  The profiles and other results in the figures shown here are
color-coded to provide a bridge between these two perspectives.

The basic idea behind our analysis is that the strength of observed 
line profiles at a given observed velocity is
inversely proportional to the LSR velocity gradient at that velocity, 
taken with respect to line of sight distance.  Multiplying an 
observed profile by that velocity gradient removes the distortions induced 
by line of sight velocity effects.  The practical use of this analysis
is to extend to individual profiles the same methods that are used in 
Galactic plane surveys to trace the variation of quantities such as the
mean densities of  H I or \HH\ across the Galactic disk.

We define R as galactocentric radius and \R0\ as the galactocentric distance
of the Sun.  Defining the line of sight distance to a point at galactocentric 
distance R at Galactic longitude $l$ as $\rho$(R) the geometry gives

$$ \rmR^2 = \rho^2+\R0^2-2\rho\R0\col \eqno(1a) $$

and 

$$ \rho = \R0\col \pm (\rmR^2-\R0^2\si2l)^{\frac{1}{2}} \eqno(1b) $$

when the velocity and longitude have the same sign, R $<$ \R0\ and there are 
two possibilities for the kinematic distance at any given velocity, or

$$ \rho = \R0\col + (\rmR^2-\R0^2\si2l)^{\frac{1}{2}} \eqno(1c) $$

when the velocity and longitude have opposite sign, R $\ge$ \R0\ and
there is no kinematic distance ambiguity for gas outside the
solar circle.

Taking the derivative in eqn 1a,

$$ \frac{d\rmR}{d\rho} = \frac{(\rho - \R0 \col)} {\rmR} \eqno(2) $$

We define $\Theta(\rmR)$ as the Galactic rotational velocity field and $\Theta_0$
= $\Theta(\rmR_0)$ as the speed of Galactic rotation at the solar circle. 
The observed (LSR) velocity induced by Galactic rotation is

$$ \rmv(\rmR) = \R0 \sil * [\frac{\Theta(\rmR)}{\rmR} - \frac{\thsun}{\R0}] \eqno(3) $$

so

$$ \frac{d\rmv}{d\rmR} = -\R0 \sil*[\frac{\Theta(\rmR)}{\rmR^2}-
 \frac{1}{\rmR} \frac{d\Theta(\rmR)}{d\rmR} ] \eqno(4)$$

Expressing the line of sight velocity gradient as

$$ \frac{d\rmv}{d\rho} = \frac{d\rmv}{d\rmR}\frac{d\rmR}{d\rho} \eqno(5)$$

and assuming a flat Galactic rotation curve $\Theta(\rmR) = \Theta_0$, 
combining eqns 2, 4 and 5 gives 

$$ \frac{d\rmv}{d\rho} = \R0 \sil * (\R0\col-\rho)* \frac{\Theta_0}{\rmR^3} \eqno(6) $$

and inverting eqn 3 to give

$$ \rmR(\rmv) = \Theta_0 * [\frac{\rmv}{\R0\sil}+\frac{\thsun}{\R0}]^{-1}\eqno(7) $$

provides a complete specification of the line of sight velocity gradient as a
function of velocity and implied galactocentric distance within the line profile.

The basic distance-velocity relationships along the line of sight to W31C are 
illustrated in Fig. 4.

\subsubsection{Random gas motions}

Random gas motions, neglected in this explicit, analytic approach,
have generally been ignored in galactic survey work because 
they should average out in large samples.  They would act to 
blur and broaden the distributions in galactocentric radius that are
derived here.  They could be incorporated by convolving the expression for
v(R) in  Eqn 3 with a gaussian having the appropriate velocity dispersion
and proceeding numerically. 

Toward W31C the velocity gradient dv/dR varies from 5 \kms\ (kpc)$^{-1}$ 
at R = 8 kpc to 22 \kms\ (kpc)$^{-1}$ at R = 4 kpc with a rough mean 
$<$dv/dR$>$ $\approx$ 50 \kms/4.5 kpc = 11 \kms\ (kpc)$^{-1}$.  This can 
be compared with the measured one-dimensional
cloud-cloud velocity dispersion of 3.0 - 4.2 \kms\ \citep{Cle85,LisBur+84}.  
The intrinsic broadening caused by random motions toward W31C is small at
R = 4 kpc, and important at larger galactocentric radii near the Solar Circle.

\subsection{Deriving the Galactic CO emission abundance}

Surveys of Galactic CO emission compute the abundance of CO emission per kpc
of path in the Galactic disk (in units of K-\kms\ kpc$^{-1}$) and
plot or tabulate this quantity as a function of galactocentric radius;  
it is variously  labelled  as \ACO\ \citep{BurGor78}, as here, or 
J$_0$ \citep{CleSan+88} or called an emissivity \citep{PinLan+13}.  
This is done by assuming a rotation curve,
and then by determining the velocities within
a profile corresponding to pre-defined bins in galactocentric radius.
The appropriately-integrated CO emission in the profile (units of K-\kms) 
and the line of sight path length in the Galactic disk are assigned to each bin 
and the final abundance of the survey
in each bin is the summed integrated emission divide by the summed path
length, as in Fig. 8 of \cite{CleSan+88} or Fig. 7 of \cite{PinLan+13}.  
The emission abundance histogram 
shows the radial  distribution of molecular gas over the Galaxy, and was the 
vehicle by which the presence of the Galactic molecular ring was first
established.  The confinement of molecular gas well inside the radius
of the H I disk was a seminal result in ISM studies.

The analogous quantity for a single CO profile is found by computing

$$ {\rm m} * \ACO(\rmv) = \Tstar (\rmv) * \abs{\frac{d\rmv}{d\rho}(\rmv)} \eqno(8) $$ 

at each velocity channel, treating the velocity v= v(R) thus giving 
\ACO (R).  In this expression, m = 2 if the gas is presumed to be
distributed with equal liklihood at the near and far kinematic distances 
and m = 1 when the gas is on only one side of the sub-central point or when 
no kinematic distance ambiguity exists.   In Galactic surveys where large
numbers of sightlines are employed, m = 2 is always assumed. 

\subsection{Deriving the  mean number density from absorption profiles}

With absorption profiles the observed optical depth of species Q at velocity
v is  proportional to the local number density n(Q) and inversely proportional 
to the Galactic velocity gradient, i.e. $\tau_{\rm Q}(\rmv)~\propto~$ n(Q)/(dv/d$\rho)$).
Hence multiplying an optical depth spectrum $\tau_{\rm Q}(\rmv)$ by a properly scaled 
velocity gradient $d\rmv/d\rho$ yields the local number density n(Q) and the
conversion from velocity
to galactocentric radius R provides the Galactic perspective.  As with the
emission the quantity determined is

$$ {\rm m} * \rmn({\rm Q}) (\rmv)\propto\tau_{\rm Q}(\rmv)*\abs{\frac{d\rmv}{d\rho}(\rmv)} \eqno(9) $$ 

with m = 2 if there is a kinematic distance ambiguity and m = 1 otherwise
(as in the present case).

Here we employ this technique to derive to derive the local space-averaged
\footnote{i.e. at the galactocentric radius corresponding to that particular velocity} 
number density $<$n(\HH)$>$ 
from CH absorption, using N(CH) $= 3.64\times10^{13}\pcc \int \tau_{\rm CH}(\rmv) d\rmv$
\citep{GerdeL+10a} and N(CH)/N(\HH) = X(CH) = $3.5\times10^{-8}$ 
\citep{SheRog+08} so that N(\HH) $= 1.04 \times
10^{21} \int \tau_{\rm CH}(\rmv) d\rmv \pcc$.  We derive the mean number density
of all H-nuclei density $<$n(H)$>$ from the C II absorption profile using 
N(C\p) $= 1.434\times10^{17}\pcc \int \tau_{\rm C II}(\rmv) d\rmv$
\citep{GerRua+15} and the free gas phase of carbon N(C\p)/N(H)
= $1.4\times10^{-4}$  \citep{SofLau+04,GerRua+15} so that N(H) $= 1.024 \times
10^{21} \int \tau_{\rm C II}(\rmv) d\rmv \pcc$.  

The microscopic {\it in situ} number densities of the medium  producing 
the observed lines differ from these space-averaged mean densities by a 
volume filling factor which is derived in Sect. 5 along with the {\it in situ}
number density and other quantities..

\subsection{The CO-\HH\ conversion factor}

The CO emission abundance  \ACO\ is useful because 
the CO-\HH\ conversion factor \XCO\ is just the ratio $<$n(\HH)$>$/\ACO, 
appropriately scaled, i.e. with 
\XCO\ = $<$n(\HH)$>$ $\times ~(3.086 \times 10^{21} {\rm cm})$/\ACO\ given that the 
units of \ACO\ are K-\kms\ kpc$^{-1}$.

\subsection{Galactic metallicity gradient}

Our analysis depends on the abundances of the carbon-bearing species CH and 
(to a lesser extent) C\p\ with respect to hydrogen, which we take to be constant 
at N(CH)/N(\HH) = $3.5\times10^{-8}$ and 
N(C\p)/N(H) = $1.4\times10^{-4}$.  \cite{GerRua+15} found no Galactic gradient in 
N(C\p)/N(H) when N(\HH) was derived from N(CH) but this could also  be the
case when the molecular fraction is high and the abundances of CH and C\p\ scale
in the same way with the local metallicity.  The discussion is further complicated
by the fact that the relevant quantities are the free gas phase abundances
after depletion, whereas the Galactic gradient is measured on the metallicity 
itself.

In their derivation of the CO-dark \HH\ gas fraction \cite{LanVel+14} used an 
exponential Galactic carbon abundance gradient N(C\p)/N(H) $\propto$ exp(-R/6.2 kpc)
following \cite{WolMcK+03}, based on the [O/H] metallicity gradient at that epoch.  
This would imply a factor two difference between the value at R = 8.5 kpc and 
that at the inner edge of the Galactic molecular ring at R = 4 kpc.  More recent 
discussions,  for instance Sect. 4.3.3 of \cite{LucLam11}, cite smaller gradients 
with much longer scale lengths, ranging from 12.5 kpc for [C/H] in Cepheids 
\citep{LucLam11} to 25 - 35 kpc for [C/H] and [O/H] in disk B-stars and H II regions 
\citep{DafCun04,MacCos10}.  Such small gradients are understandable in terms
of radial mixing, see \cite{KubPra+15} and they imply  much smaller variations:
For a scale length of 25 kpc, the change in metallicity over the range 
R = 4 - 8.5 kpc is 20\%.  On this basis the Galactic metallicity 
gradient is not included in the baseline C/H and CH/\HH\ abundance ratios used
in this work.
 
\begin{figure}
\includegraphics[height=7.75cm]{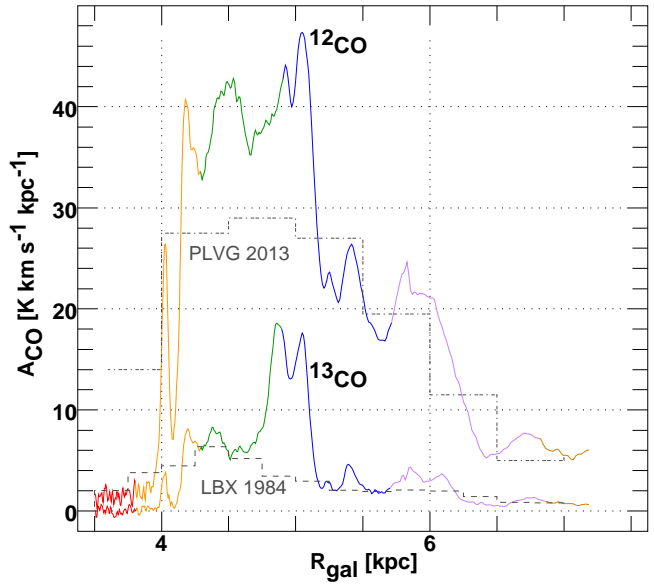}
  \caption[]{Galactic abundance of CO and \coth\ emission vs. 
galactocentric distance for a flat rotation curve with R$_0$ = 8.5 kpc and
$\Theta_0$ = 220 \kms.  The dashed  histogram of the \coth\ abundance labelled 
LBX 1984 is from a re-analysis of the \coth\ survey data of 
\cite{LisBur+84} and the dashdot histogram for CO labelled PLVG 2013 was transcribed 
from Fig. 7 of  \cite{PinLan+13}.  The color coding is as in previous figures.}
\end{figure}

\section{Gas and cloud properties along the line of sight to W31C}

Here we follow the methods  outlined in Sect. 4, using the line profiles
toward W31 to derive the CO emission abundance, the mean H I and \HH\ 
densities and the quotient of the mean molecular density and CO emission 
abundance that is the CO-\HH\ conversion factor.
We derive {\it in situ} number densities, the typical volume
filling factors of atomic and molecular gas and the size scale of
clumping in the molecular gas.

\subsection{The CO emission abundance}

Fig. 5 shows the emission abundances for CO and \coth\, assuming (as in
the case of blind surveys) that the gas is equally distributed at the 
near and far distances (m = 2 in eqn 8).  Our curves in Fig. 5 would be 
twice higher for gas assumed to lie on only one side of the Galaxy.

Comparison with published Galactic survey results provides a valuable 
perspective on the line of sight to W31, but the derived quantities from 
older Galactic CO surveys that observed the Galaxy most widely were 
computed for \R0\ = 10 kpc, $\Theta_0$ = 225 \kms\ and have never been 
re-cast.  They scale as \ACO\ $\propto$ \R0$^{-1}$ in only the most approximate 
sense.  Nonetheless we see from Fig. 8 of \cite{CleSan+88} a peak 
${\rm J}_0 * 10/8.5 \approx 30$ K-\kms kpc$^{-1}$,
 compared with 40 K-\kms\ kpc$^{-1}$ in our Fig. 5.  

A modern version of the 
CO emission abundance histogram using data in the Galactic plane is shown in 
Fig. 7 of \cite{PinLan+13}, from which we transcribed the trace shown 
in Fig. 5 here.  The larger-scale emission abundance histogram of \cite{PinLan+13}
coincides with the results derived toward  W31 for R $\ga$ 5 kpc
but has a somewhat smaller peak value 28 K \kms\ kpc$^{-1}$ at smaller galactocentric
radii.

Fig. 5 compares  the abundance histogram for \coth\ emission toward W31 with that 
observed in the Galactic equator at $l$ = 20\degr\ - 40\degr\ by 
\cite{LisBur+84}, recomputed for the flat rotation curve employed here.   
In this case the two peaks at 28 - 30 \kms\ toward W31 stand out.  Otherwise, the 
larger-scale survey results generally coincide with the result toward W31.

\subsection{Locally space-averaged atomic, molecular and total densities}

Fig. 6 at top shows the overall Galactic distribution of the locally
space-averaged  atomic and molecular mid-plane densities drawn from 
\cite{PinLan+13}.   The analogous space-averaged mean atomic and molecular
densities along the line of sight to W31C after applying the 
analysis in Sect. 4 to the H I, CH and C II profiles are shown in the middle panel.
The Galactic molecular ring structure seen in CO and C\p\  emission  is clearly 
present toward W31 but the space-averaged atomic and  molecular number densities 
in the molecular ring toward W31 are 5-6 times higher than the Galactic
average.  

To derive the H I gas properties we averaged profiles from the 
Southern Galactic Plane Survey H I (SGPS) server {\footnote 
http://www.atnf.csiro.au/research/HI/sgps/} on either side
of W31 at the same latitude and converted the emission to optical depth 
assuming \TK\ = 135 K.  We then applied eqn 9 as discussed in Sect. 4.3 
for the velocity interval 10 - 50 \kms\ with the result that 
N(H I) $= 1.36 \times 10^{22} \pcc$.
The distance to W31 is 4.95 kpc \citep{SanRei+14} and the line 
of sight distance to a position at R = 6.82 kpc where v = 10 \kms\ is 1.73 kpc,
so that the mean space-averaged H I number density is 1.37 $\pccc$. 
This is some three times larger than the Galactic average $<$n(H I)$> \approx 0.45 \pccc$
shown in Fig. 6.   As also shown in Fig. 6 the space averaged H I number 
density toward W31C varies over the range $1 \pccc \la ~<$n(H I)$> ~\la 2 \pccc$.

The integrated optical depth of the CH absorption profile above 10 \kms\ is 
18.81 \kms\ implying N(CH) $= 3.64\times10^{13}\pcc \int \tau({\rm CH}) d\rmv$
= $6.85\times10^{14}\pcc$ and N(\HH) = N(CH)/$3.5\times 10^{-8} = 1.96 \times
10^{22} \pcc$. Therefore the mean density of H-nucleii in \HH\ is 
2$<$n(\HH)$>$ = 2 N(\HH)/3.23 kpc = 3.95$\pccc$.  
This is about two times higher than the Galactic mean at the peak of the 
Galactic ring in Fig. 6 but only half the peak toward W31.   As shown
in Fig. 6 the space averaged \HH\ number density varies over the range
$2 \pccc \la 2<{\rm n}(\HH)> ~\la 12\pccc$, much more widely than for n(H I).

Also shown in the middle panel of Fig. 6 is an estimate of the total hydrogen
density n(H) toward W31 assuming a fixed carbon abundance N(C\p)/N(H) 
= $1.4 \times 10^{-4}$ as discussed in Sect. 4.  The overall agreement 
between the derived \HH\ and total hydrogen distributions is proof that 
the high molecular densities are not artifacts of a mistaken CH abundance.  
However  the C II line  generally misses the narrow peaks in the molecular 
density (which are seen in  CO and C I emission), suggesting that 
its abundance is dimininished in some portions of the absorbing material.

The integrated optical depth of the C II line above 10 \kms\ is 45 \kms\ implying
N(C\p) $= 6.5 \times 10^{18} \pcc$,    N(H) = $4.6 \times 10^{22}$ and
$<$n(H)$> = 4.6\pccc$.  There is a 6\% overall correction for the contribution
of N(C I) $\approx 4 \pm 1 \times 10^{17}\pcc$ \citep{GerRua+15}, 
i.e. N(H) = $4.9 \times 10^{22}$ and $<$n(H)$> = 4.9\pccc$ in total overall.   
N(H) derived from carbon is only very slightly smaller than the sum 
N(H I) + 2N(\HH) $= 5.28\times10^{22} \pcc$ derived from H I and CH separately.
This near-agreement implies that C II absorption does not have a
substantial contribution from WIM even though WIM probably occupies nearly
all of the volume toward W31C, as noted in Sect. 5.5 below.  The fractional
contribution of WIM toward W31C is no more than 5\%, based on the 
detection of N II absorption \cite{PerGer+14} leading to 
N(N\p)/N(C\p) = 1/40, i.e. some twenty times smaller than the [N]/[C]
elemental abundance ratio in the gas.
 
\subsection{The CO-\HH\ conversion factor}

The CO-\HH\ conversion factor is computed from the ratio $<$n(\HH)$>$/\ACO, 
properly normalized and with an assumption about the near-far
distribution of the CO emission on either side of the Galaxy.  Fig. 6 at bottom
shows the CO-\HH\ conversion factor assuming that the CO emission coincides 
with the absorbing material on the near side of W31.   Material on the near 
side of the Galaxy is much closer to the nominal
midplane, typically at $|$z$|$ $\approx$ 20 - 30 pc, well within one scale-height,
compared to material at the far kinematic distance which would be at 
${\rm z}| \approx 80 - 100$ pc.  From Fig. 4 we see that material at v = 30 \kms\
is seen at $\rho$ = 4 and 13  kpc , or at $|$z$|$ = 28 or 91 pc, compared to a scale
height for the molecular gas of 50 pc \citep{CleSan+88}.  We noted in Sect. 3 that
the strongest \coth\ emission is about symmetrically distributed in latitude with
respect to W31C, and that the gas at 12-22 \kms\ appears more nearly confined
at the equator.

The derived CO-\HH\ conversion factor shown in the bottom panel of Fig. 6 has
structure, but it varies far less at R = 4 - 6.5 kpc than either the molecular
density or the CO abundance \ACO\ (shown as a faint trace in the lower panel
of Fig. 6).  The value derived here in the inner Galactic disk derived by 
generalizing the  CH chemistry in the vicinity of the Sun, 
\ACO\ $= 1.5\pm0.5 \times 10^{20}~ \HH~ \pcc$ (K \kms)$^{-1}$ is comparable to the 
canonical value $ 2 \times 10^{20}~ \HH~ \pcc$ (K \kms)$^{-1}$ that is used
in Galactic studies.  


\begin{figure}
\includegraphics[height=10cm]{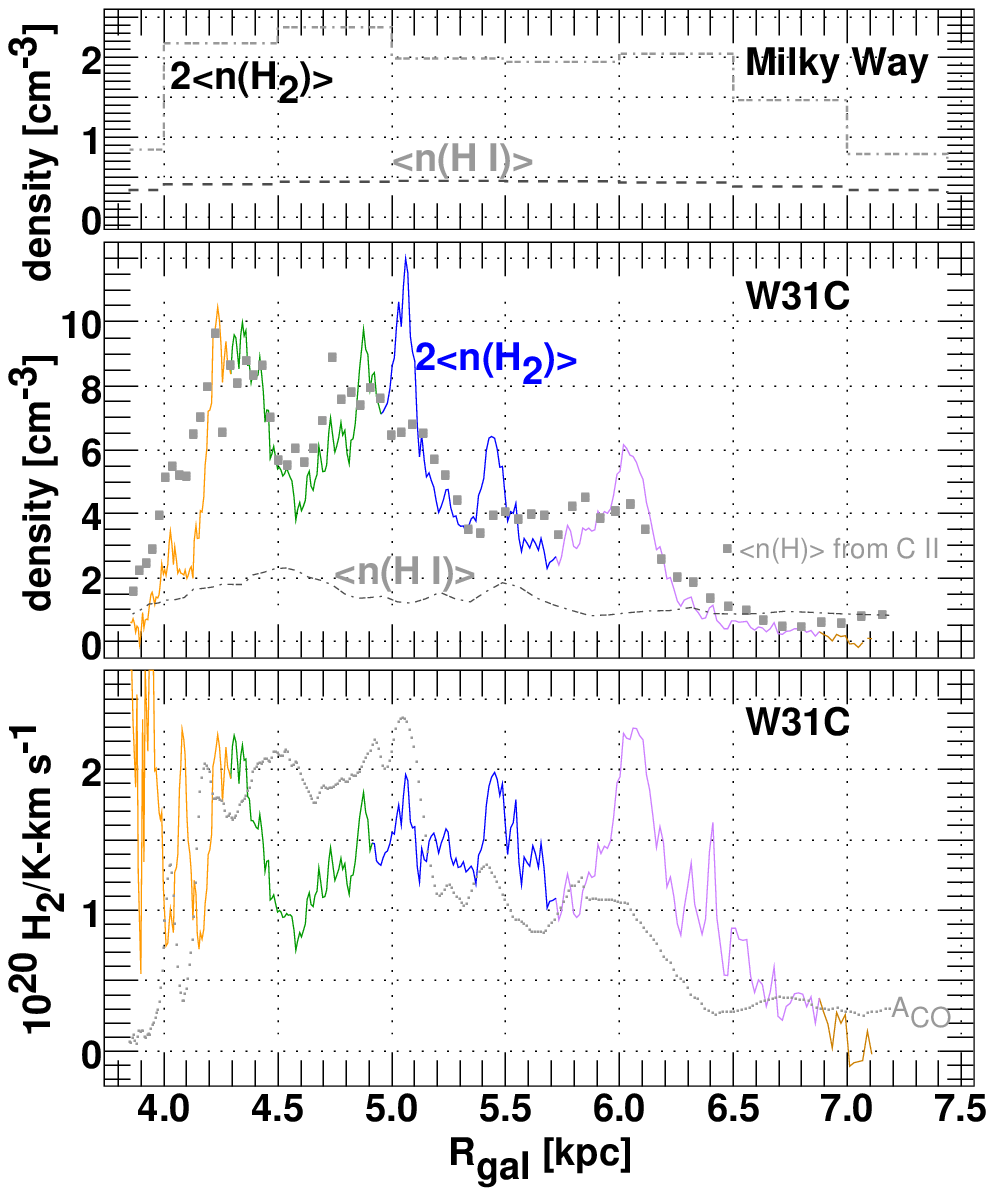}
  \caption[]{Top: Milky Way space-averaged mean atomic and molecular 
 densities from \cite{PinLan+13}.
  Middle: mean densities 2$<$n(\HH)$>$ derived from CH absorption with 
  X(CH) = $3.5\times 10^{-8}$ shown as a solid line, color-coded as 
  in previous figures); $<$n(H)$>$ derived from C II absorption using n(C\p)/n(H)
  = $1.4\times 10^{-4}$ (filled light gray rectangles); and n(H I)
  derived from from a nearby NGPS survey H I emission profile assuming a constant 
 spin temperature of 135 K, shown as a dashed grey line.  Bottom: 
 the CO-\HH\ conversion factor derived by dividing the mean molecular 
 density $<$n(\HH)$>$ by the CO abundance \ACO\  assuming 
  that the CO-emitting gas is on the near side . The histogram of \ACO\ is 
 overlaid and heavily shaded with arbitrary vertical scale in the bottom 
 panel.}
\end{figure}

\subsection{{\it In situ} number density}

\subsubsection{The {\it in situ} number  density of \HH-bearing gas}

The easiest and widest surveys of  molecular gas in the Galactic disk are 
those of CO emission, which is notoriously insensitive to the density.   
In diffuse gas when CO is weakly-excited and far below thermalization, 
the excitation temperature of the J=1-0 line of CO is 
proportional to the thermal pressure of \HH, corrected for the line optical 
depth \citep{SmiSte+78,LisLuc98,LisPet12}.  A similar situation arises with the fine
structure lines of neutral atomic carbon \citep{JenTri11} from which the 
thermal presssure of the diffuse ISM (but not the temperature or density
separately) is usually derived.

Comparison of the emission brightness and absorption optical depth of 
strongly-polar species such as  \hcop\ can be used to derive the {\it in situ} number 
density to which emission or absorption measurements alone
are notoriously insensitive.  A minimum number density estimate can be derived 
under the assumption that the gas is diffuse, with most of the free gas phase 
carbon in the form of C\p\ so that electrons provide most of the rotational 
excitation for strongly-polar species \citep{Lis12HiMu}.  Because the
excitation arises from electrons that are contributed by C\p\ and atomic 
hydrogen the density so derived is the total density n(H) and 
is only weakly dependent on the molecular fraction.   The mean brightness 
of the \hcop\ emission shown in Fig. 1, 0.03 K and the mean ratio of the 
brightness of \hcop\ to CO, 1\%, are entirely typical of what 
is observed for diffuse cloud sightlines near the Sun \citep{LucLis96}, 
but observations of C\p, CH and CO toward W31 provide constraints on this 
interpretation, so that it need not be accepted ad hoc.


\begin{figure}
\includegraphics[height=11cm]{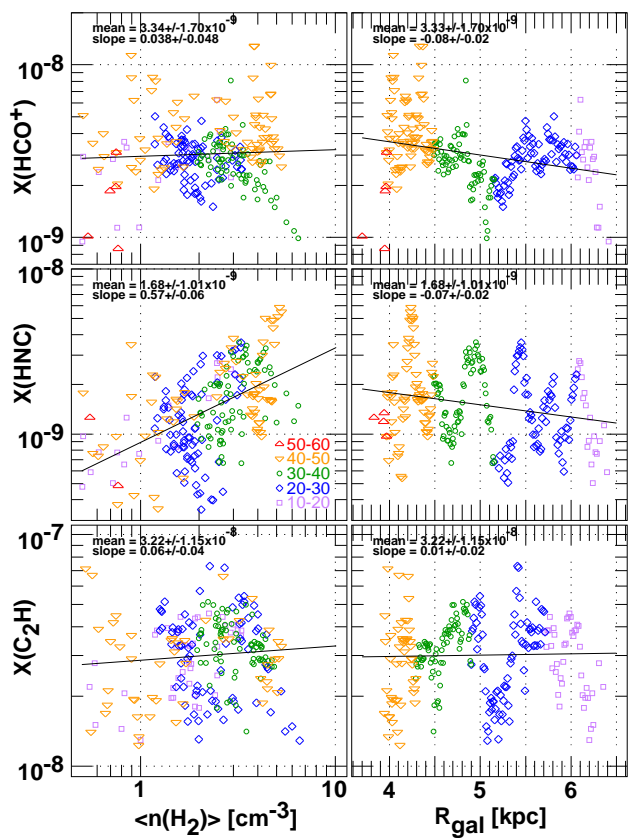}
  \caption[]{Relative abundances of \hcop, HNC and \cch\ observed
in absorption with respect to \HH\ plotted against the local
 spaced-averaged mean \HH\ density derived from CH (at left) and 
galactocentric radius (at right).  Column and number densities
of \HH\ were derived from the CH absorption profile as discussed in Sect. 
4 of the text.   The mean relative abundances $\pm 1\sigma$ and 
power-law slopes of the fitted regression lines are shown in the 
individual panels.}
\end{figure}

Fig. 1 of \cite{Lis12HiMu} plots the predicted brightness of the 
J=1-0 line of \hcop\ vs dN(\hcop)/dV (which we derive from the optical
depth spectrum):  Results are parametrized with the number density at 
a fixed electron fraction x$_e$ = n(e)/n(H) = $1.4\times10^{-4}$ as
assumed here for the relative abundance of C\p, and with the electron 
abundance at 
n(H) = 256 $\pccc$.  For convenience, Fig. B.1 here shows dN/dV for \hcop,
HNC and HCN across their absorption spectra and indicates their means, 
which for \hcop\ is $<{\rm dN/dV}> = 1.5 \times 10^{12}\pcc$ (\kms)$^{-1}$.  
The integrated brightness of the \hcop\ emission profile shown in Fig. 1  
over the velocity range from 10 to 50 \kms\ is 1.16 K-\kms\ and the 
mean brightness over the range 10 $\le$ v $\le 50$ \kms\ is 0.029 K.  

From the left hand panels of Fig. 1 in \cite{Lis12HiMu} we find that 
the mean \hcop\ brightness is reproduced at the mean column density
per unit velocity  if 
n(H) $\approx 160 \pccc$ for x$_e = 1.4\times10^{-4}$.  Because it 
ignores other contributions to the electron fraction that causes 
most of the excitation of \hcop, this is an upper limit.
At this density C\p\ should contribute at least 80\% of the total electron 
abundance \citep{GerRua+15} with most of the remainder from H\p\ and a 
smaller contribution from He\p, implying n(H) $\approx 0.8*160 =
128 \pccc$.  However, the electron fraction contributed by C\p\ 
decreases with density as the gas supports a higher level of 
cosmic-ray ionization, allowing the density to be still somewhat
smaller.   We estimate $110 \pccc \la $ n(H) $\la 160 \pccc$ 
and use n(H) $\approx 135\pm 25\pccc$ in the discussion below
as a coarse average.

\subsubsection{The {\it in situ} number density of atomic gas}

The {\it in situ} number densities and kinetic temperatures measured in C II by 
\cite{GerRua+15} toward W31 were n(H) = $40-45 \pccc$ and \TK\ = 90 - 105 K,
presumably a weighted average of the molecular component at 
n(H) $\approx 135 \pccc$ and a less dense, somewhat warmer atomic component.
 The thermal pressure toward W 31 fell in a narrow range about 
p/k = n \TK\ $= 4000 \pccc $ K, implying n(H I) $\approx 30 \pccc$ at
the temperature \TK\ = 135 K that we used to derive the mean H I density
in Fig. 6 from the observed 21cm emission.

\subsection{H I and \HH\ volume filling factors}

Because the space-averaged densities $<$n(H I)$> = 1.36 \pccc$ and  
2$<$n(\HH)$> = 3.95 \pccc$ are so much smaller than the {\it in situ}
densities n(H) $\approx 30 \pccc$ and n(H)$ \approx 135 \pccc$ in
the predominantly atomic and molecular components respectively,
the neutral interstellar medium along the sightline to W31 must be 
highly clumped, with overall line of sight average volume filling factors 
\fvol(H I) $\approx 1.36\pccc/30\pccc = 4.6$\% 
and \fvol(\HH)  $\approx 3.95\pccc/135\pccc = 3$\%,
each varying over the range 3-7\% at constant {\it in situ} density.  

\cite{GerRua+15} found an overall Galactic median total volume filling 
factor \fvol\ = 2.4\%\ based on their interpretation of C II absorption 
toward the wide sample of 
sightlines in the Galactic plane sampled in the PRISMAS project, but 
with much higher filling factors 6\%, 13\% and 8\% over
the velocity intervals 10-23 \kms, 23-34 \kms and 34 - 61 \kms\ toward W31. in
overall agreement with the total filling factors derived here from
other considerations.

The comparably small volume filling factors determined separately from 
C\p\ \citep{GerRua+15} and the combination of $<$n(H I)$>$ and $<$n(\HH)$>$ 
have several implications.  That they are so much less than one implies 
that some 90\% of volume along the line of sight to W31C must be occupied 
by ionized gas, either warm (WIM at 8000 K) or hot ($10^6$ K).  That they 
are comparable implies that most of the C II absorption arises in neutral 
gas, whether cold or warm, as opposed to WIM, a conclusion that is
reinforced by the near equality of the total column densities N(H) 
inferred separately from carbon and from the sum of H I and scaled CH
as discussed in Sect. 5.2

\begin{figure}
\includegraphics[height=13.8cm]{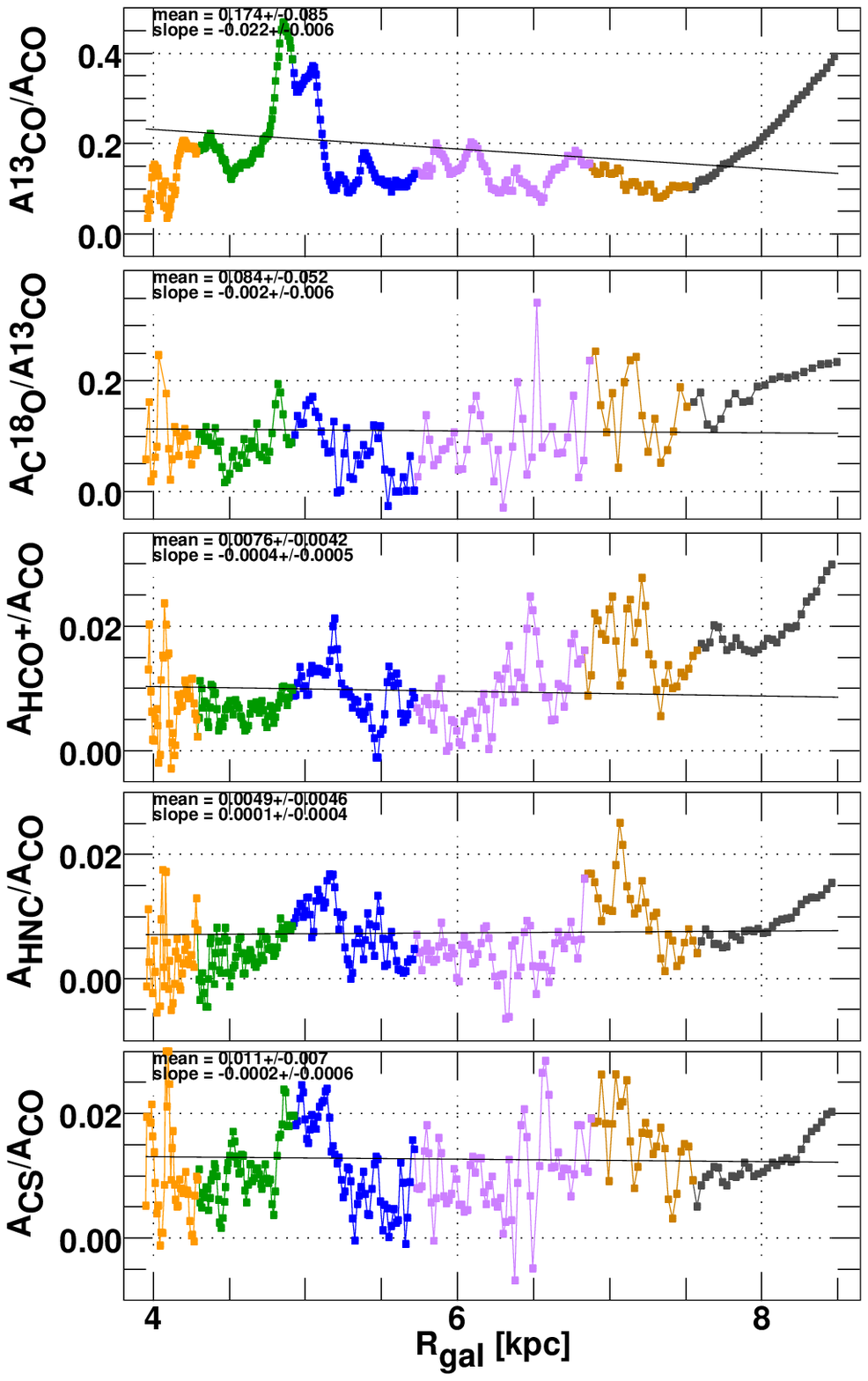}
  \caption[]{Ratios of emission from \coth, \hcop, HNC,
and CS(2-1) relative to CO and \coei\ relative to \coth.  The regression lines were 
fit to the data at R $\le$ 6.6 kpc (R $\le$ 6.3 kpc for \hcop) to avoid 
contamination from the W31 complex  at V $\le$ 10 \kms\ that appears as a 
false excess of emission at larger galactocentric radii for species with higher dipole moments.  
 The mean ratios $\pm 1\sigma$ and slopes  of the
fitted regression lines are shown in the individual panels.}
\end{figure}


\subsection{A cloud size scale for clumping in the \HH-bearing gas}

A size scale for cloud clumping may be derived from the continuous nature
of the emission and absorption over the range v = 10 - 50 \kms.
With $<$ dN(\hcop)/dV $> = 1.5 \times 10^{12} \pcc$ (\kms)$^{-1}$ and X(\hcop)
$= 3 \times 10^{-9}$ (Fig. 7 or \cite{LisPet+10}) the implied mean 
column density of clouds with linewidth dV (in units of \kms)  is 
N$_{\rm cld}$(H) $\approx$ 2N(\HH) $= 10^{21}\pcc$ dV
\footnote{i.e. N(\HH) = (dV $\times$ dN(\hcop)/dV)/X(\hcop)}.  
For dV = 2-4 \kms\ these would be classic diffuse-translucent clouds with 
\AV\ $\approx$ 1-2 mag.

The number of clouds along the line of sight \#$_{\rm clds}$ must be at least
what is required to span the 40 \kms\ profile width, or 
\#$_{\rm clds} \ga 40 ~\kms$/dV.  The mean free path of the clouds along 
the 3.23 kpc path corresponding to the velocity range 10-50 \kms\  
is then $\Lambda \la 3.23$ kpc/\#$_{\rm clds} \la$ 0.081 dV kpc and 
the typical cloud size scale D can be inferred from the derived volume 
filling factor \fvol\ $\approx$ D/$\Lambda \approx 0.03$ implying D $\la$ 2.7 dV pc.  
In turn the mean internal number density of such clouds n$_{\rm cld}$(H) = 
N$_{\rm cld}$(H)/D $\ga 120 \pccc$, independent of dV.  Thus the parameters 
inferred from the \hcop\ chemistry and excitation separately yield
a  consistent picture 
of the host cloud  medium.  A cloud with linewidth dV = 2 \kms\ would have 
D = 5.4 pc, 
which at the mean line of sight distance to W31, 2.5 kpc, would subtend an 
angle of 7.4\arcmin\ = 0.124\degr, see Fig. 3.   

\subsection{CO, \coth\ and \coei\ emission}

To be sure, CO emission from the Galactic plane is not normally interpreted
as arising in diffuse/translucent clouds of density n(H) $\approx 135 \pccc$
so we must inquire whether the brightnesses of the carbon monoxide lines are 
also consistent with this interpretation.  The integrated CO brightness 
\WCO\ = 146 K \kms\ at v = 10 - 50 \kms\ requires 
N(CO) $\approx 1.5\times 10^{17}\pcc$ at the 
typical brightness of diffuse molecular gas where 1 K \kms\ corresponds to 
N(CO) =  $10^{15} \pcc$ \citep{Lis07CO}.   This is somewhat smaller
than the column density of neutral atomic carbon
N(C I) $\approx 4 \pm 1 \times 10^{17}\pcc$ and very small compared to 
the integrated C\p\ column density N(C\p) = $6.5 \times 10^{18}$ derived 
in Sect 5.2 from the observed C II absorption line profile of \cite{GerRua+15}.  
The main isotope CO emission is indeed compatible with the small number density 
derived from \hcop.

The mean \coei/\coth\ brightness ratio, shown in Table 1 and Fig. 8, is 
$0.084\pm0.052$ compared to the isotopic abudance ratio 65/560 $\approx$ 
0.12 near the Sun or perhaps a slightly smaller value 50/560 $\approx$ 
0.09 expected in the inner galaxy \citep{MilSav+05}.  Thus the \coth\ 
emission nominally appears to be optically thin, despite the relatively
high \coth/CO brightness ratio $0.18\pm0.09$ (Table 1 and Fig. 8) or
where the ratio is as high as 0.4. 

CO/\coth\ brightness ratios as small as 5 are typically interpreted 
in terms of thermalized and very optically thick CO emission from dense 
cold molecular clouds which is inappropriate for the line of sight toward 
W31.  Substantial fractionation of $^{13}$C is expected for the gas toward 
W31 given that N(CO) $<<$ N(C\p) but this has never been taken into account
for Galactic CO emission.  

\section{Cloud chemistry and and Galactic gradients}
 
\subsection{Variations of relative abundances with $<$n(\HH)$>$ and R}

Fig. 7 shows the relative abundances derived from the absorption spectra of \hcop,
HNC and \cch\ \citep{GodFal+10}, again using the CH absorption profile as a surrogate
for  \HH.  The nearly-fixed relative abundances 
$<$X(\hcop)$> = 3.3 \pm 1.9 \times 10^{-9}$ and $<$X(\cch)$> = 3.2\pm1.2 \times10^{-8}$
derived in this way with respect to $<$n(\HH)$>$ are the same as those derived 
in previous absorption line work at high latitude in the Solar neighborhood 
\citep{LucLis00C2H,LisPet+10} or in the Galactic plane more generally \citep{GerKaz+11}.  
The relative abundances of \cch\ and CH are very nearly the same.

The sharper variation of X(HNC) with respect to $<$n(\HH)$>$ at left in Fig. 7
with correlation coefficient r = 0.52 and power-law slope $= 0.57\pm0.06$ 
can be interpreted as implying that regions of higher mean \HH\ density, 
generally at smaller galactocentric radii (see Fig. 6), have a larger contribution 
from material having a higher mean N(\HH) per cloud, because absorption 
observations against compact 
continuum sources at high latitude near the Sun have shown that the column densities 
of the CN-bearing species HNC, HCN, and CN all increase abruptly when 
N(\hcop) $\ga 10^{12} \pcc$ \citep{LisLuc01}.

A variation in mean cloud column density could also be responsible for the 
small but statistically significant radial gradient in X(HNC) with 
galactocentric radius  (power-law slope $= -0.074\pm0.022$) 
seen at the right in Fig. 7: the correlation coefficient, is only r=0.22.  
However, the same explanation does not obviously account for the comparable Galactic 
gradient in X(\hcop), (power-law slope $= -0.077\pm0.016$ 
and correlation coefficient r=0.29) if X(\hcop) does not vary with 
$<$n(\HH)$>$  at left in Fig. 7.

%
%

\subsection{Variations relative to CO}

Table 1 gives integrated intensities and ratios with respect to \WCO\ for the 
pointed observations shown in Figs. 1, 2 and A.1 and  Fig. 8 shows the 
radial variations in the ratios \coth/CO, \coth/\coei, 
 \hcop/CO, HNC/CO and CS(2-1)/CO from the same emission spectra.  Regression 
analysis was carried out for the channels corresponding to 
 R $<$ 6.6 kpc (R $<$ 6.3 for \hcop)  at v $\ga$ 10-15 \kms, so as to avoid
contamination by the red wing of the emission profile from W31, whose
effects are evident at larger radii. 

The only statistically significant Galactic gradient in Fig. 8, amounting to a 
factor two across the Galactic disk, is in the \coth/CO ratio.  A comparable 
gradient seen over the longitude range l = 20\degr - 40\degr\ was previously 
ascribed to a decrease in the mean column N(H) per cloud at larger radii 
\citep{LisBur+84} although some role might be reserved for variation in 
the $^{13}$C/$^{12}$C abundance ratio \citep{MilSav+05}.  A variation
in mean cloud column density was already suggested from the variation of
X(HNC) with $<$n(\HH)$>$ in the immediately previous subsection here.  
The \coei/\coth\ ratio does not vary detectably, implying that optical depth 
effects in \coth\ across the Galactic disk are small, but also that a 
Galactic gradient in the $^{13}$C/$^{12}$C ratio is also not an important 
factor in variation of the \coth/CO ratio.

No statistically significant variations with galactocentric radius are detectable
in the \hcop/CO, HNC/CO or CS(2-1)/CO ratios.  This  follows the  pattern of 
previous work where such variations were undetected in wider-scale surveys 
along the Galactic equator in the inner Galaxy \citep{Lis93,Lis95,HelBli97}. 
This general lack of variation seems surprising given the changes that are
expected across the Galactic disk \citep{WolMcK+03}, the change in mean molecular
density that is apparent in Fig. 6, and the implied change in the mean cloud
column density that was inferred from the strong variation in the CO/\coth\ ratio.
Moreover, X(HNC) appears to increase with $<$n(\HH)$>$ in Fig. 7 and
  $<$n(\HH)$>$ is larger at smaller galactocentric radii in Fig. 6.
Apparently, comparison with \WCO\ somehow compensates for other changes,
perhaps in the same manner that the CO-\HH\ conversion factor also does not change
systematically over the inner galaxy because CO emission and CH absorption so
closely resemble each other.  \cite{SakHas+97} found a decreasing 
CO J=2-1/J=1-0 ratio at R $>$ 6.5 kpc for gas lying out of the Galactic plane, but
variations in the CO J=2-1/J=1-0 line ratio are not apparent for gas in the
Galactic plane \citep{Lis93,SakHas+97}.

\section{Summary}

This paper is a follow-on to the PRISMAS project on Herschel, which studied 
emission and absorption from C\p\ and absorption from a wide variety of molecular 
hydrides, hydride ions and heavier molecules at mm and sub-mm wavelengths 
toward H II regions used as  background light ources in the inner galaxy.  As described in 
Sect. 2 and 3, we  presented emission profiles of the J=1-0 lines of CO, \coth, \coei, 
\hcop\ and HNC and the J=2-1 line of CS taken at the older ARO 12m Kitt Peak antenna 
toward and around the Galactic HII region W31C G10.62-0.38 at 1\arcmin\ resolution.
Comparison of  molecular emission and absorption spectra provides estimates of
the {\it in situ} number density and volume filling factors of the host gas 
that can not be obtained from emission or absorption separately.  From a 
comparison of CO emission and CH absorption we gave an independent, 
chemically-based derivation of the CO-\HH\ conversion factor.

The sightline to W31C G10.62-0.38 at a line of sight distance $\rho$ = 4.95 kpc 
is the most strongly absorbed of all those that were observed with PRISMAS in the 
inner Galactic disk.  CH and \hcop\ absorption that best trace \HH\ are continuous 
with relatively little contrast over the profile (see Figs. 1 and 2).  \cotw\ and 
\hcop\ emission profiles also show little differentiation into kinematically distinct 
cloud features. Even HNC, which preferentially appears in the most strongly-molecular
regions, has continuous absorption and emission across  the full velocity range over 
which \coth\ emission occurs (Fig. A.1).  

Although the LSR velocity of W31C (-3 \kms) is anomalous given its location
4.95 from the Sun at galactocentric radius R = 3.9 kpc just inside the 
molecular ring, the kinematics of the intervening path are well characterized 
by Galactic rotation.  The velocity interval over which absorption is present, 
up to 50 \kms\, corresponds to that expected from Galactic rotation.  
Velocity-distance transformations for the line of sight to W31C are shown in Fig. 4,
based on the formalism developed in Section 4.  
Given the continuous and somewhat amorphous nature of the absorption and emission, 
we developed a method of kinematic analysis relating line profile velocity to 
line of sight distance $\rho$ and galactocentric radius R via the line of sight 
velocity gradient for a flat Galactic rotation curve with R$_0$ = 8.5 kpc and 
$\Theta_0$ = 225 \kms.  The formalism is described in Sect. 4 where  we showed how 
optical depth is related to locally space-averaged number density in the Galaxy
(denoted here by $<$n(H I)$>$, $<$n(\HH)$>$ etc) 
allowing a derivation of $<$n(\HH)$>$ with 
galactocentric radius from CH absorption assuming a constant fractional abundance of 
CH with respect to \HH, X(CH) = N(CH)/N(\HH)$ =3.5\times 10^{-8}$.  We also showed 
how to derive the  emission abundance 
\ACO\ (units of K-\kms\ kpc$^{-1}$) that historically was derived statistically 
from large-scale CO surveys and used, for instance, to discover the Galactic 
molecular ring.  The ratio $<$n(\HH)$>$/\ACO, properly scaled, is the CO-\HH\ 
conversion factor.

The emission abundances of CO and \coth\ were discussed in Sect. 5.1 and
shown in Fig. 5 where they are compared with previous results
from Galactic surveys.  The  emission abundances toward W31C show the familiar
Galactic molecular ring at R = 4-6 kpc, with an added contribution interior to
R = 5 kpc. 

In Sect. 5.2 (see Fig. 6) we characterized the sightline to W31C 
as having locally space-averaged atomic and molecular densities 
$<$n(H I)$>$ $\approx 1-2 \pccc$ and 2$<$n(\HH)$>$ $\approx 2-10 \pccc$ 
galactocentric radii 4 \kpc $<$ R $<$ 6.5 kpc with overall means 
$<$n(H I)$> = 1.36 \pccc$ and $<$n(\HH)$> = 3.95 \pccc$. These may be 
compared to  $<$n(H I)$> \approx 0.45 \pccc$ and 2$<$n(\HH)$>\approx 2 \pccc$  
for the galaxy at large at the same galactocentric radii.  The high mean 
densities toward W31C presumably arise because the Sagittarius, Scutum-
Centaurus and Norma spiral arms all run across the line of sight.

We derived the space-averaged \HH\ density $<$n(\HH)$>$ 
from the CH absorption profile and the relative abundance 
N(CH)/N(\HH) $= 3.5\times 10^{-8}$ observed in the ISM near the Sun 
and derived the CO-\HH\ conversion factor 
$<$n(\HH)$>$/\ACO\ $= 1.5\pm0.5\times 10^{20}$ \HH\ (K-\kms)$^{-1}$ over 
the inner galaxy, close to the canonical value 
N(\HH)/\WCO\ $= 2 \times 10^{20}$ \HH\ (K-\kms)$^{-1}$ (see Sect. 5.3
and the bottom panel in Fig 6).  The CH absorption
and CO emission profiles resemble each other sufficiently that the
ratio $<$n(\HH)$>$/\ACO\ varies much less than either  $<$n(\HH)$>$ or \ACO\ alone.
This supports our assumption that the CO emission arises in front of W31C, 
which we argued was likely to be the case given the large Galactic z-height
at the far kinematic distance.

The {\it in situ} number density of the molecular gas was derived in Sect. 
5.4.1 by comparing emission and absorption in \hcop\ and found to be modest, 
2n(\HH)  $\approx 135 \pm 25 \pccc$.   The H I column density toward W31C was 
derived assuming a kinetic temperature of 135 K and the {\it in situ} 
number density of the atomic gas at this temperature is n(H I) = $30 \pccc$ 
at the thermal pressure p/k $= 4000 \pccc$ K derived previously from C\p\ 
(Sect 5.4.2). The mean H I and \HH\ densities  bracket
 the density derived earlier from C II, n(H) $\approx 60 \pccc$.   
The CO column density N(CO) $ = 1.5 \times 10^{17} \pcc$ inferred at 
2n(\HH) $= 135 \pccc$ is small compared to the column density of neutral
atomic carbon N(C I) $ = 4 \times 10^{17} \pcc$ and very small compared 
to the total carbon column density N(C) $= 7 \times 10^{18}\pcc$ derived
from observations of the C II and C I lines.

Comparison of the space-averaged and {\it in situ} densities toward W31C 
in Sect. 5.5 translate to volume filling factors 3-7\% for H I or \HH, in 
agreement with the overall volume filling factors 8-13\% derived earlier from 
C II absorption that arises ubiquitously throughout all but the very densest 
neutral gas.  The density contrast and volume filling factor of the
molecular gas imply clumping on a scale of 5.5 pc or 7\arcmin\ at the mean 
distance to W31C as noted in Sect 5.6.

As discussed in Sect. 6.1, Fig. 7 at left shows the variation of relative 
abundance with $<$n(\HH)$>$ 
for \hcop, HNC, and \cch\ representing three distinct chemical families.
The relative abundances X(\hcop) = N(\hcop)/N(\HH) 
$= 3.3 \times 10^{-9} $ and X(\cch) $= 3.2 \times 10^{-8} $ toward W31C are 
equal to values observed in the ISM near the Sun.  Given that we used 
CH to infer $<$n(\HH)$>$, it follows that the \hcop/CH and \cch/CH ratios 
do not vary across the  Galactic disk.  

The \hcop\ and hydrocarbon relative abundances are nearly fixed in the chemistry 
observed near the Sun so it is probably not surprising that they show little 
variation with $<$n(\HH)$>$ at left in Fig. 7.  The increase of X(HNC) with 
$<$n(\HH)$>$ is perhaps understandable in terms of the  rapid increase of the 
column densities of the CN-bearing
species in local diffuse clouds when N(\hcop) $\ga 10^{12} \pcc$:
Regions of higher $<$n(\HH)$>$ could have a larger contribution from clouds with
higher column density.  The decline in X(HNC) with galactocentric radius shown
at right in Fig. 7 would then imply that clouds observed at larger R have smaller 
 column density on average, which has previously been invoked to explain the gradient 
in \coth/CO that is shown in Fig. 8.  However, the decline in X(\hcop) with
galactocentric radius is not understandable in these terms. 
  
In Fig. 8 we showed the ratio of emission brightness of various species relative to
CO across the Galactic disk and found variation only in the \coth/CO ratio, 
from 0.2 at R= 4 kpc to 0.1 at R = 8.5 kpc,
a decline by factor two over the Galactic disk interior to the Sun as previously 
seen in CO surveys over the inner galaxy.  As noted in Sect. 6.2, this was previously
interpreted as implying that the mean cloud column density is smaller at
larger galactocentric radii, consistent with the behaviour of HNC discussed
in the previous paragraph.

The ratios of brightnesses \hcop/CO, HNC/CO and CS(2-1)/CO, all $\approx$ 1\%, 
do not vary over the inner Galactic disk in Fig. 8 and, as noted in  Sect 6, this 
is consistent with wider but still  limited surveys in the Galactic plane comparing  
emission from \hcop, CS, HCN with CO and \coth.   Those surveys showed that cloud
features seen in \coth\ are also seen in the the rarer species with higher
dipole moments at levels 1\% - 2\% that of CO.   The ubiquitous emission and
absorption from species other than CO toward W31C is an example of the same phenomenon.

\subsection{recap}

There is a very substantial legacy of absorption spectra from the Herschel
mission.  A relatively small investment of time spent taking complementary 
molecular emission spectra could be used to derive a wealth of additional 
information from that data.


\begin{appendix}

\section{Emission spectra of HCO$^+$, HNC and CS J=2-1}

Shown in Figure A.1 are the scaled emission spectra of all species except \cotw,  
including the profile of  CS J=1-1 that is not shown elsewhere in this work.
A comparison of all these species with CO is shown in Fig. 8.

\begin{figure}
\includegraphics[height=7.15cm]{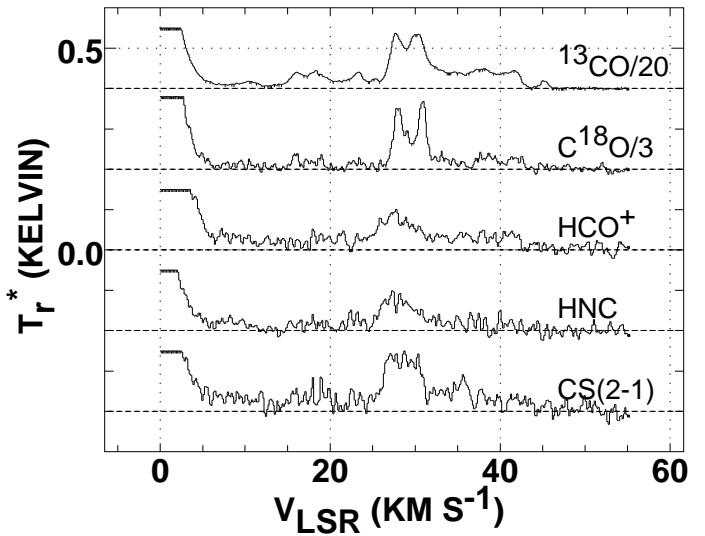}
  \caption[]{Weaker emission profiles observed at the ARO, including the HNC and 
CS J=2-1 lines that are not shown elsewhere in this work.}
\end{figure}

\section{Column density spectra}

Shown in Fig. B.1 are the column density spectra dN/dV of HCN, HNC and \hcop\ derived
from their absorption profiles.  Mean values $<$dN/dV$>$ are indicated 
at left in the figure and used in Sect. 5.5.1 to derive the {\it in situ} hydrogen 
densities n(\HH) from a comparision of \hcop\ emission and absorption.

\begin{figure}
\includegraphics[height=7.25cm]{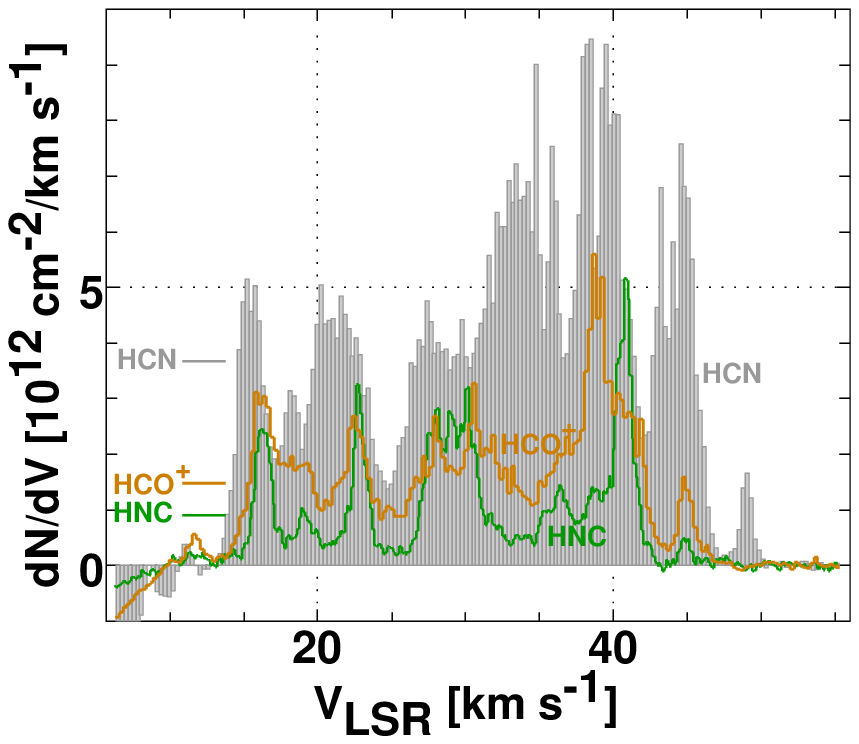}
  \caption[]{Optical depth spectra converted to column density per unit velocity.  Indicated at
left are the mean values at 10 \kms $ \le \rmv \le 50$ \kms,  
$<$ dN(HCN)/dV $> = 3.66 \times 10^{12} \pcc$ (\kms)$^{-1}$,
$<$ dN(\hcop)/dV $> = 1.48 \times 10^{12} \pcc$ (\kms)$^{-1}$ and
$<$ dN(HNC)/dV $> = 0.89 \times 10^{12} \pcc$ (\kms)$^{-1}$
}
\end{figure}

%
%

\end{appendix}

\begin{acknowledgements}
The National Radio Astronomy Observatory is a facility of the National 
Science Foundation operated under cooperative agreement by Associated 
Universities, Inc. The Arizona Radio
Observatory is operated by Steward Observatory, University of Arizona, with partial
support through the NSF University Radio Observatories program (URO: AST-1140030).
We are grateful to the TAC and the Director of ARO for the opportunity to observe
at the Kitt Peak 12m Telescope. This work was partly funded by grant 
ANR-09-BLAN-0231-01 from the French 
{\it Agence Nationale de la Recherche} as part of the SCHISM project.
We thank the referee for comments that led us to clarify the text in
several regards.

\end{acknowledgements}
 
\bibliographystyle{apj}


\end{document}